\documentclass[preprint]{aastex}
\usepackage{natbib}
\usepackage{pdflscape}
\def\apjs{ApJS}

\def\apjl{ApJ}
\def\aap{A\&A}

\def\asec{\ifmmode ^{\prime\prime}\else$^{\prime\prime}$\fi}

\def\it{\sl}
\def\degs{\ifmmode ^{\circ}\else$^{\circ}$\fi}
\def\amin{\ifmmode ^{\prime}\else$^{\prime}$\fi}
\def\asec{\ifmmode ^{\prime\prime}\else$^{\prime\prime}$\fi}
\def\fm{\hbox{$.\!\!^{\rm m}$}}            % Fractions of magnitudes
\def\fdg{\hbox{$.\!\!^\circ$}}          % Fractions of degrees
\def\farcs{\hbox{$.\!\!^{\prime\prime}$}}  % Fractions of arcseconds
\def\psr{PSR~J2021+3651}
\def\degs{\ifmmode ^{\circ}\else$^{\circ}$\fi}
\def\amin{\ifmmode ^{\prime}\else$^{\prime}$\fi}
\def\farcm{\hbox{$.\mkern-4mu^\prime$}}
\def\eqalign#1{\null\,\vcenter{\openup1\jot \m@th
   \ialign{\strut\hfil$\displaystyle{##}$&$\displaystyle{{}##}$\hfil
   \crcr#1\crcr}}\,}

\slugcomment{submitted to Astrophysical J.}

\shorttitle{Deep optical observations of PSR J2021$+$3651
}
\shortauthors{Kirichenko et al.}
\begin{document}

%% LaTeX will automatically break titles if they run longer than
%% one line. However, you may use \\ to force a line break if
%% you desire.

\title{Optical observations of PSR J2021+3651 in the Dragonfly Nebula 
with the GTC\thanks{Based on observations made with the Gran Telescopio Canarias (GTC),
instaled in the Spanish Observatorio del Roque de los Muchachos
of the Instituto de Astrofísica de Canarias, in the island of La Palma, programme GTC3-11B.
}
}

%% Use \author, \affil, and the \and command to format
%% author and affiliation information.
%% Note that \email has replaced the old \authoremail command
%% from AASTeX v4.0. You can use \email to mark an email address
%% anywhere in the paper, not just in the front matter.
%% As in the title, use \\ to force line breaks.
\author{Aida Kirichenko\altaffilmark{1,2}, Andrey Danilenko\altaffilmark{1}, Peter Shternin\altaffilmark{1,2}, 
Yuriy Shibanov\altaffilmark{1,2}, Elizaveta~Ryspaeva\altaffilmark{2}, Dima Zyuzin\altaffilmark{1}, 
Martin Durant\altaffilmark{3},
Oleg Kargaltsev\altaffilmark{4}, 
George Pavlov\altaffilmark{5},\\ and Antonio Cabrera-Lavers\altaffilmark{6,7}}
\altaffiltext{1}{Ioffe Institute, Politekhnicheskaya  
 26, St.~Petersburg, 194021, Russia}
\altaffiltext{2}{St.~Petersburg  Polytechnic University, 
Politekhnicheskaya 29, St.~Petersburg, 195251, Russia} 
\altaffiltext{3}{Department of Medical Biophysics, Sunnybrook 
Hospital M6 623, 2075 Bayview Avenue, Toronto M4N 3M5, Canada} 
\altaffiltext{4}{Department of Physics, The George Washington University, Washington, DC 20052, USA}
\altaffiltext{5}{Department of Astronomy \& Astrophysics, Pennsylvania State University, 525 Davey Lab,
University Park, PA 16802, USA} 
\altaffiltext{6}{Instituto de Astrofísica de Canarias, E-38205 La Laguna, Tenerife, Spain}
\altaffiltext{7}{Universidad de La Laguna, Dept. Astrofísica, E-38206 La Laguna, Tenerife, Spain}

%% Notice that each of these authors has alternate affiliations, which
%% are identified by the \altaffilmark after each name.  Specify alternate
%% affiliation information with \altaffiltext, with one command per each
%% affiliation.

%% Mark off your abstract in the ``abstract'' environment. In the manuscript
%% style, abstract will output a Received/Accepted line after the
%% title and affiliation information. No date will appear since the author
%% does not have this information. The dates will be filled in by the
%% editorial office after submission.

\begin{abstract}  

 \psr\ is a 17 kyr old rotation powered pulsar detected in the radio, X-rays, and $\gamma$-rays. 
 It powers a torus-like  pulsar wind nebula with jets, dubbed the Dragonfly, which is 
 very similar to that of the Vela pulsar. The Dragonfly is likely associated with 
 the extended TeV source VER J2019+368 and extended radio emission. 
 We conducted first deep optical observations with the GTC in the Sloan $r'$ band  
 to search for  optical counterparts of the pulsar and its nebula. No counterparts were detected down 
 to $r'\ga27.2$ and $\ga24.8$ for the point-like pulsar and the 
 compact X-ray nebula, respectively. We also reanalyzed \textit{Chandra} archival X-ray data taking into account an interstellar extinction 
 -- distance relation,
 constructed by us for the Dragonfly line of sight using the red-clump stars as standard candles.
 This allowed us to constrain the distance to the pulsar, $D=1.8^{+1.7}_{-1.4}$ kpc at 90\% confidence.
 It is much smaller than the dispersion measure distance of $\sim$12 kpc 
 but compatible with a $\gamma$-ray ``pseudo-distance'' of 1 kpc. Based on that and the optical upper limits, 
 we conclude that \psr, similar to the Vela pulsar, is a very inefficient nonthermal emitter in the optical and 
 X-rays, while its $\gamma$-ray efficiency is consistent with an average efficiency for   $\gamma$-pulsars of similar age.               
 Our optical flux upper limit for the pulsar is consistent with the long-wavelength extrapolation of 
 its X-ray spectrum while  the nebula flux upper limit does not constrain the respective extrapolation. 

\end{abstract}


\section{Introduction} 
\label{sec:intro}
  
The 104 ms pulsar J2021+3651~was discovered in the radio with the Arecibo telescope in 
a deep search for radio pulsations towards  unidentified
\textit{ASCA} X-ray sources spatially coinciding with \textit{EGRET} 
$\gamma$-ray objects \citep{roberts2002ApJ}. 
With the characteristic age $\tau_c$~$\approx$ 17 kyr and spin-down luminosity 
$\dot{E}$~$\approx$ 3.4~$\times$~10$^{36}$ erg~s$^{-1}$ 
this pulsar is among the youngest and most energetic rotation-powered pulsars known. 
The  canonical dipole magnetic field estimated from the pulsar 
period and its derivative is $B$~$\approx$ 3.2~$\times$~10$^{12}$ G. 
A tentative detection of $\gamma$-ray pulsations with the pulsar period  
in the \textit{EGRET} data  was reported by \citet{2003McL-cordes}.  
Later $\gamma$-ray observations with \textit{AGILE} satellite  
\citep{halpern2008ApJ} and \textit{Fermi} observatory \citep{abdo2009ApJ} 
firmly  established a  double-peaked pulse 
profile and  a power-law spectrum  with a photon index $\Gamma \sim 
1.7$ and cutoff energy of 
$\sim$~2.9 GeV \citep{2013Abdo}. \psr\ was also identified in X-rays with \textit{Chandra}, and weak pulsations, also with 
the double-peaked profile, were detected at a 4$\sigma$ 
significance \citep{hessels2004ApJ,abdo2009ApJ}. The pulsar X-ray spectrum contains thermal 
and non-thermal components from the surface and magnetosphere of a neutron star (NS), 
respectively \citep{vanetten2008ApJ}. 
\textit{Chandra} also revealed an extended pulsar wind nebula (PWN) G75.2+0.1 whose 
brightest internal part, within $\sim$~30\asec~of the pulsar, has a torus-like morphology 
with axial jets. By its specific spatial shape 
this PWN was dubbed the Dragonfly Nebula \citep{vanetten2008ApJ}. 
A fainter diffuse emission 
is extended up to  several arcminutes.  

The \psr\ position is projected on the Cygnus-X region, one of 
the richest known regions of star formation in the Galaxy.  
A bright  extended TeV source MGRO J2019+37 was identified 
with the \textit{Milagro} sky survey in this region 
with a 20 TeV flux of 80\% of those of the Crab Nebula \citep{abdo2007ApJ}.
The source was suggested to be 
associated with the Dragonfly,  
which was recently confirmed by observations with the VERITAS 
observatory. VERITAS resolved the source into two objects \citep{aliu2014ApJ}.  
The brightest one, VER J2019+368, has a hard 
spectrum resembling the spectrum of Vela X -- a TeV PWN system powered by the 
Vela pulsar. VER J2019+368 coincides also 
with an extended region of non-thermal radio emission.     

The most controversial parameter of 
\psr\ is the distance.
The NE2001 model for the Galactic distribution of free electrons \citep{cordes2002Arxiv} for 
the pulsar line of 
sight ($l$~= 75\fdg21, $b$~= 0\fdg11) and the dispersion measure  $DM$~$\approx$ 370 pc~cm$^{-3}$
yield a distance  $D$~$\approx$ 12 kpc \citep[e.g.,][]{roberts2002ApJ}. 
Comparing a hydrogen absorbing column density obtained from first X-ray observations and the total Galactic 
HI column density along the pulsar line of sight, \citet{hessels2004ApJ} suggested $D\approx 10$ kpc.   
\citet{vanetten2008ApJ} performed similar analysis of subsequent deeper X-ray observations and found 
the distance of 3--4~kpc.
The pulsar polarization rotation measure 
implies a minimal $D$~$\approx$ 5 kpc \citep{abdo2009ApJ}. 
Adopting the latter value and assuming that \psr\ was born near 
the center of VER J2019+368, \citet{aliu2014ApJ} estimated a possible transverse 
velocity of the pulsar to be $\sim$~840 km s$^{-1}$, 
which is about 3 times higher than the average for pulsar velocities \citep{hobbs}. 
Finally, the distance can be as low as  
1.5 kpc if the pulsar is located within the Cygnus-X region. 
This location is consistent with the
empirical $\gamma$-ray ``pseudo-distance'' relation \citep[e.g.,][]{saz2010ApJ} 
suggesting $D\sim1$ kpc.

By many multiwavelength properties  
\psr\ and its PWN are similar to the 
Vela pulsar plus PWN system, but, in contrast to the Vela, this pulsar has 
never  been studied in the optical.   
We report first deep optical observations of the \psr\ field performed with 
the 10.4-m Gran Telescopio Canarias (GTC). We also address the issue of the distance discrepancies, using the 
\textit{Chandra} archival X-ray 
data and red-clump stars as standard candles, and compare the optical results with the X-ray ones. 
The details of  observations and data reduction are described
in Sect.~\ref{sec:opt}, our results are presented 
in Sect.~\ref{sec:res} and are discussed
in Sect.~\ref{sec:dis}.
%%%%%%%%%%%%%%%%%%%%%%%%%%%%%%%%%%%%%%%%%%%%%%%%
\begin{figure*}[t]
\setlength{\unitlength}{1mm}
\begin{center}
\begin{picture}(160,125)(0,0)
\put (0,55) {\includegraphics[width=77.0mm, bb= 84 200 500 562 clip=]{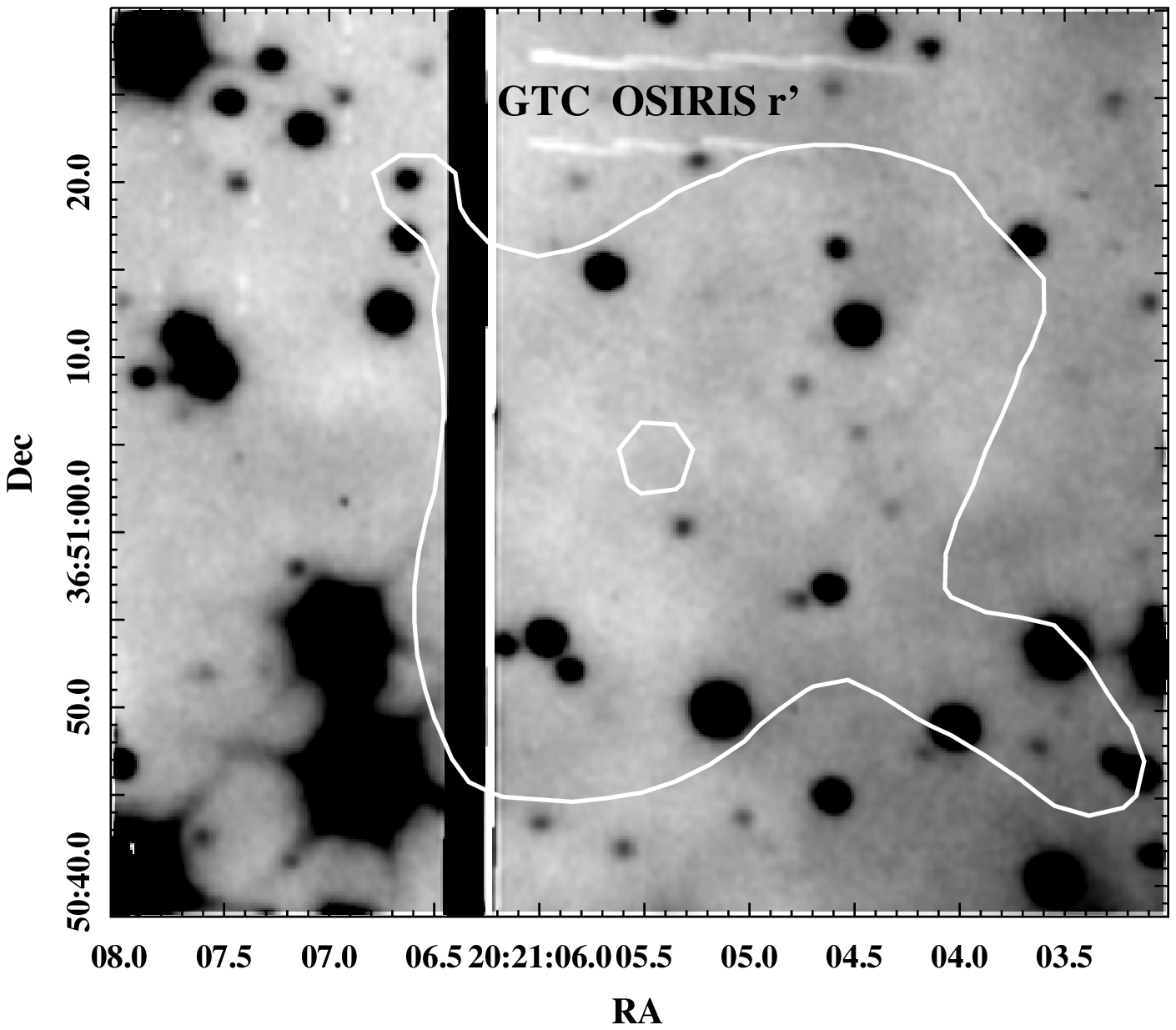}}
\put (80,55) {\includegraphics[width=77.0mm,  bb= 84 200 500 562 clip=]{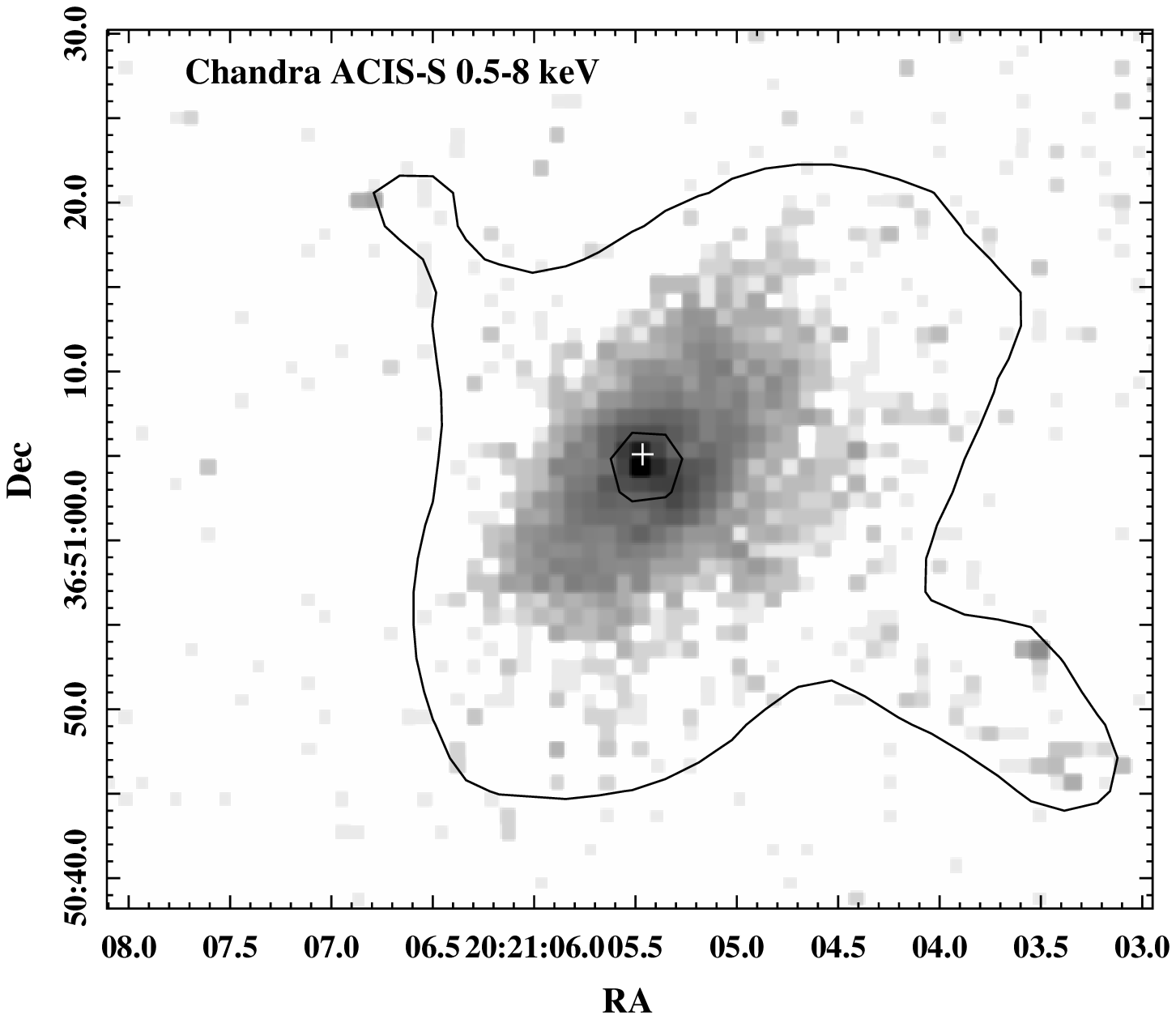}}
\put (4,-17) {\includegraphics[width=66mm, bb=99 191 486 561 clip=]{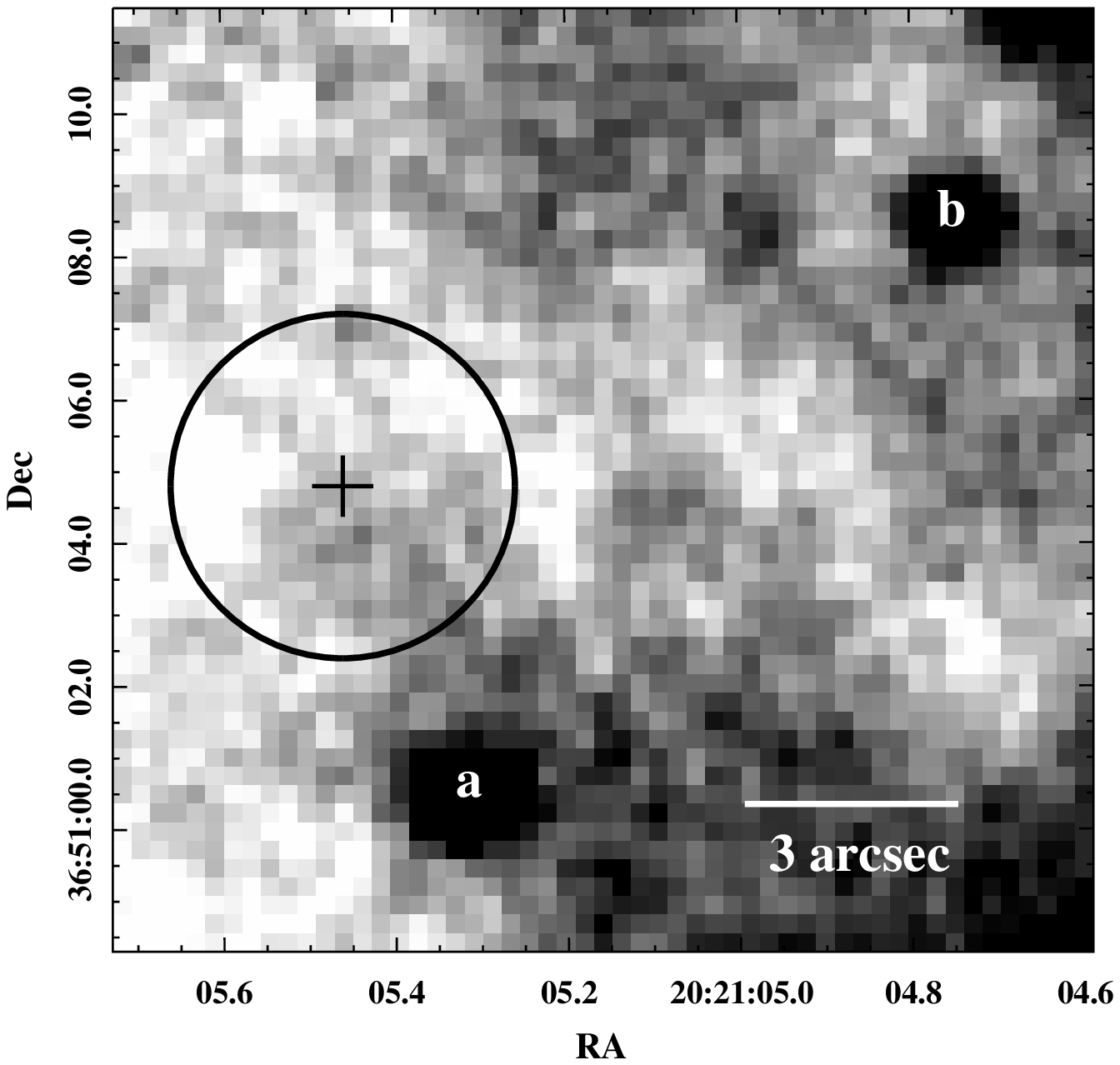}}
\end{picture}
\end{center}
\vspace{-57mm}
\hfill\parbox[b]{80mm}{\caption{{\sl Top}: $\sim$~60\asec~$\times$~60\asec~GTC/OSIRIS optical Sloan 
$r'$ ({\sl top-left}) and \textit{Chandra}/ACIS-S X-ray ({\sl top-right})
 image fragments of the \psr\ field. The pulsar  is marked by ``$+$'' in the X-ray 
 image and  contours indicating the X-ray PWN boundary and pulsar location region are shown in both images. 
 {\sl Bottom}: 13\asec~$\times$~13\asec~fragment of the Sloan $r'$-band depicting the pulsar vicinity. 
 The circle shows 3$\sigma$ X-ray pulsar
 position uncertainty. Two  sources nearest to the pulsar are labeled. }}
\vspace{10mm}
\label{fig:ima}
\end{figure*}
%%%%%%%%%%%%%%%%%%%%%%%%%%%%%%%%%%%%%%%%%%%%%%%%%%%%

\section{GTC data}
\label{sec:opt}
%%%%%%%%%%%%%%%%%%%%%%%%%%% OBSERVATIONS
\subsection{Observations and data reduction}
%%%%%%%%%%%%%%%%%%%%%%%%%%%%
\label{sec:obs}
The pulsar field was imaged with the Optical System for
Imaging and low-intermediate Resolution Integrated Spectroscopy
(OSIRIS\footnote{For instrument details see {http://www.gtc.iac.es/instruments/osiris/}}) in the Sloan $r'$ band 
at the  GTC on 2011 September 28. Sixteen dithered 158-second exposures were obtained using 
the OSIRIS standard image scale of 0\farcs254/pixel with the field of view 
of 7\farcm8~$\times$~7\farcm8.  
The field was exposed on a mosaic of two CCDs, with the target source placed on CCD2. 
The observations were carried out during dark time, and the conditions were 
photometric, with seeing varying from 0\farcs8 to 1\farcs1. 

Standard data reduction and analysis, including bias subtraction and flat-fielding,  
was performed with {\tt IRAF} tools. To eliminate  
shifts  between  individual exposures, we collected a set of unsaturated stars in the field 
and aligned the images to the one with the best seeing using {\tt IRAF} routines. 
The alignment uncertainty was $\la$~0.1 pixel. 
All exposures were then combined and yielded a final image with  
a mean seeing of 0\farcs9, airmass of 1.8, and total integration time of $\approx$~2.5 ks.

\subsection{Astrometric referencing and photometric calibration}
%%%%%%%%%%%%%%%%%%%%%%%%%%%%%%%%%%%%%%%%%%%%%
\label{sec:astroref}

For astrometric referencing, positions of 10 suitable astrometric standards from the 
USNO-B1.0 catalogue\footnote{http://www.nofs.navy.mil/data/fchpix/} were used. 
Their pixel coordinates were measured on the combined image with the IRAF task 
{\sl imcenter}. The IRAF {\sl ccmap} routine was 
applied to the astrometric transformation of the image. 
Formal {\sl rms} uncertainties of the astrometric fit for the combined image were $\Delta$RA~$\la$ 0\farcs17
and $\Delta$Dec~$\la$ 0\farcs13. Accounting for the nominal catalog uncertainty
of $\approx$~0\farcs2, this results in conservative estimates of 1$\sigma$ 
referencing uncertainties of 
$\la$~0\farcs26 for RA and $\la$~0\farcs24 for Dec. 

The photometric calibration was carried out with G158-100 Sloan standard 
observed the same night as our target. The atmospheric extinction of $0.10 \pm 0.01$ mag~airmass$^{-1}$ for the Sloan $r'$-band 
was taken from the OSIRIS user manual.   
The resulting magnitude zero-point for our $r'$ image is $29\fm13\pm 0\fm02$.

\section{Results}

\label{sec:res}
\subsection{Searching for the Dragonfly and pulsar optical counterparts}
\label{sec:opt_res}

The 60\asec~$\times$~60\asec~pulsar field fragment of the GTC/OSIRIS 
$r'$-band image, which contains the brightest part of the Dragonfly 
Nebula, is shown in the top-left panel of Figure~\ref{fig:ima}. The image is smoothed with 
a Gaussian kernel with width $\sigma=2$~pix. 
It is compared with the respective \textit{Chandra}/ACIS-S 0.5--8 keV X-ray image (top-right panel 
of Figure~\ref{fig:ima}), obtained by merging all the available  archival data\footnote{PI Roberts, \textit{Chandra}/ACIS-S, OBsID 3901, 
date of obs. 2003-02-12,  19 ks exposure; PI Romani, \textit{Chandra}/ACIS-S, OBsIDs 7603 and 8502, date of obs. 2006-12-29 
59+34 ks exposure}  (112 ks effective exposure in total). 
The data were reprocessed with the CIAO v.4.6 \textit{chandra\_repro} tool with CALDB v.4.5.9. 
The X-ray image is binned by 
two ACIS pixels, smoothed with one-pixel Gaussian kernel, and shown in log-intensity scale.
The X-ray PWN is comprised of a SW jet, NE counter-jet, 
and two arcs,  which 
are oriented perpendicular to the jets. The arcs are believed to be  associated with the PWN equatorial 
torus seen almost edge on \citep{hessels2004ApJ,vanetten2008ApJ}.  
Contours in the X-ray image 
indicate the outer boundary of the torus-like PWN, where it blends with the background, 
and  the region around the pulsar (marked by the cross). 
In the top left panel of Figure~\ref{fig:ima} the X-ray contours  are overlaid on the optical image. 
The vertical bold strip crossing the left side of 
the optical image  is a bleed line from a bright over-saturated background star located outside the fragment.
Two wave-shaped horizontal curves near the top side of the image are detector artifacts.  
Comparing the optical 
and X-ray images, we do not find any significant extended optical feature correlated with the X-ray morphology of
the compact Dragonfly PWN. 
However, at an arcminute  scale comparable to the faint diffuse X-ray emission extent,  
there are some  background variations containing bright and dark regions seen in all individual 
exposures as well. Examination of H$\alpha$ images of the field from the INT Photometric Survey of the Northern Sky 
\citep{2014H-alpha} shows that these variations correlate with  the H$\alpha$ emission variations. 

The immediate pulsar vicinity is enlarged in the left-bottom panel of Figure~\ref{fig:ima}, where   
the $r\approx$2\farcs4 circle is centered at the pulsar X-ray position with 
RA~= 20:21:05.46 and Dec~= +36:51:04.8 
 \citep{hessels2004ApJ}. It corresponds to the 3$\sigma$ pulsar position uncertainty which  
accounts for the optical astrometric referencing and pulsar X-ray position uncertainties. 
No significant point-like objects are detected within the pulsar error circle. 
The closest reliably detected point-like source  ``a'' 
with $r'= 24.40 \pm 0.04$ is located 
at about 4\farcs8 or at $\approx$ 6$\sigma$ from the pulsar X-ray position. 
More distant  object ``b'' has a lower brightness of 25\fm01 $\pm$ 0\fm05 and is located 
at $\approx$ 9\farcs4 or about 12$\sigma$  from the pulsar. Because of their large  
offsets, both  sources 
are unrelated to the pulsar. 

Using our optical image we, therefore, can  set only upper limits on the pulsar and the Dragonfly Nebula flux densities in 
the Sloan $r'$ band. For the pulsar, we used a mean background deviation 
within a circular aperture with a 4 pixel ($\approx1\asec$) radius centered at the pulsar position. 
We accounted for an aperture correction of 0\fm1 derived 
using bright background stars. 
The resulting 3$\sigma$ upper limit on the pulsar flux density is $\la$0.04 $\mu$Jy ($r'\ga$27.2).
For the nebula, we used an elliptical aperture with semi-axes  of 6\farcs2 and 10\farcs6 and a position angle
of 137\degs~centered at the pulsar, which encapsulates most of the X-ray PWN equatorial torus emission. 
The 3$\sigma$ upper limit on the spatially integrated flux density of the PWN is $\la$0.36 $\mu$Jy ($r'\ga$~24.8).

%%%%%%%%%%%%%%%%%%%%% dist_diag  %%%%%%%%%%%%%%%%%%%%%%%%%%%%%%%%%%%%%%%
\begin{figure}
\setlength{\unitlength}{1mm}
\begin{center}
\begin{picture}(160,145)(0,0)
\put (20,80) {\includegraphics[scale=0.49, clip]{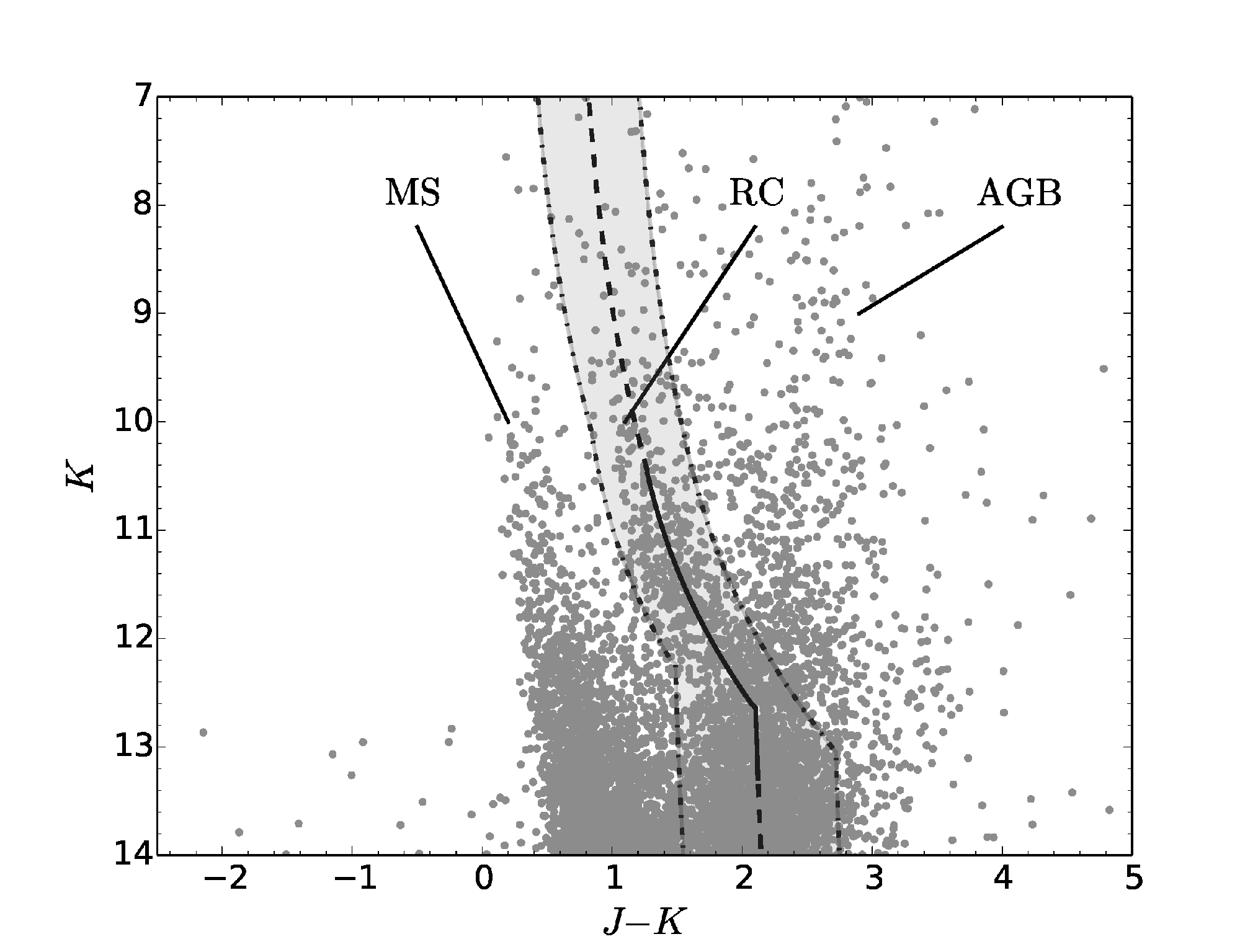}}
\put (20,0)   {\includegraphics[scale=0.49, clip]{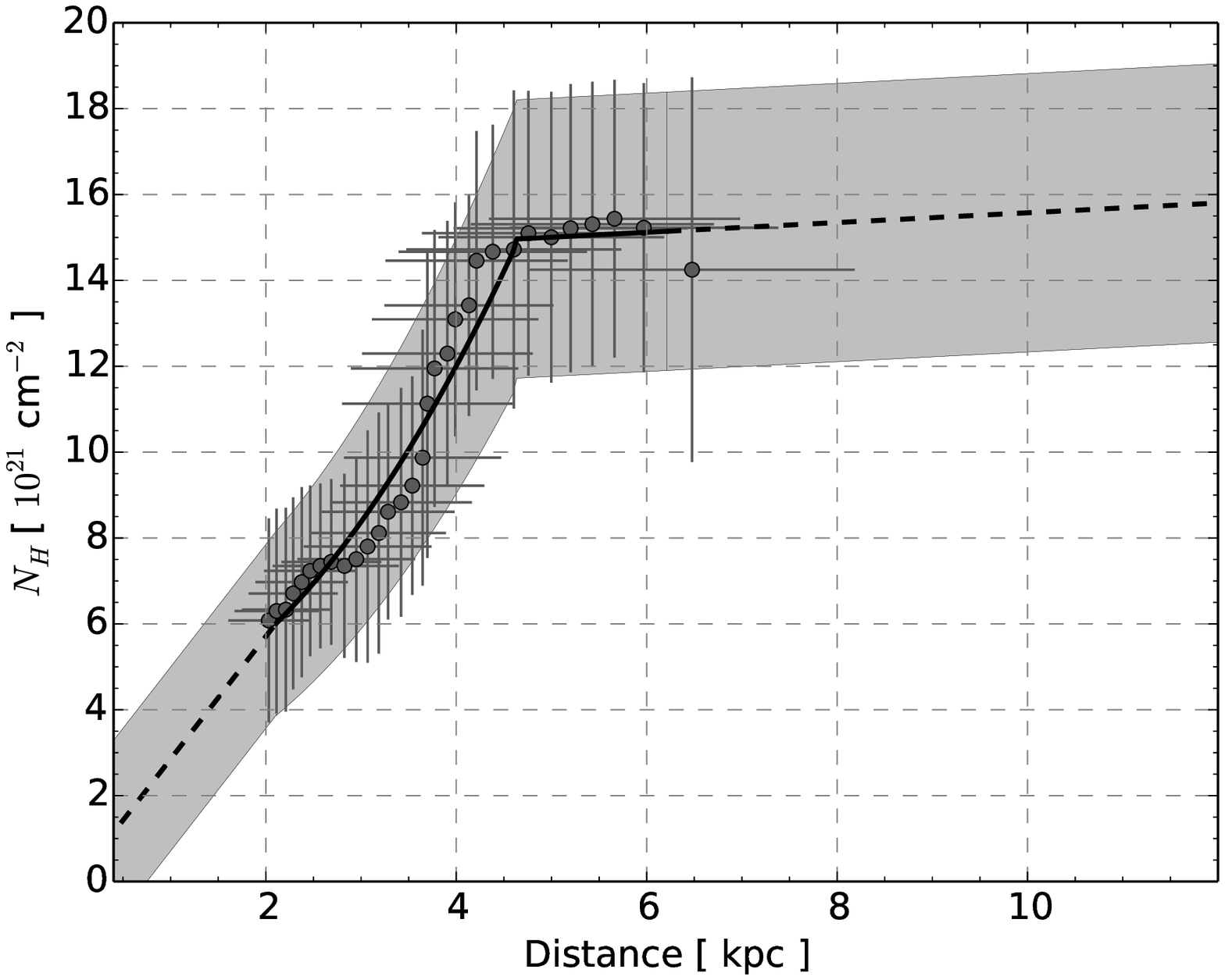}}
\end{picture}
\end{center}
 \caption{
 {\sl Top}: $K$ {\sl vs} $J-K$ diagram for the stars from the 2MASS All-Sky Point Source Catalogue 
 located within 0\fdg3 of the pulsar position ($l$ = 75\fdg22,  $b$ = 0\fdg11).
 The main-sequence (MS), red-clump  (RC), and asymptotic-giant  (AGB) branches are indicated. 
 Solid line shows a smoothed spline approximation to the RC stars mean colors; 
 dashed segments are its extrapolations to higher and lower $K$ magnitudes.
 Light-shaded region bounded by dot-dashed lines contains 95\% (2$\sigma$) of RC stars.
 {\sl Bottom}: Empirical $N_{\rm H}$--distance relation for the \psr\ direction derived using the RC stars 
 colors from the  diagram at the {\sl top} panel. Bars are 1$\sigma$ uncertainties.  
 The solid line and gray filled region are smoothing spline approximations to the data points   
 and their uncertainties, respectively. They are linearly extrapolated to small and large distances (dashed lines). 
 }
 \label{fig:nh_dist}
\end{figure}
%%%%%%%%%%%%%%%%%%%%%%%%%%%%%%%%%%%%%%%%%%%%%%%%%%%%%%%%%%%%%

\subsection{Distance and interstellar extinction}
\label{sec:dist_ext}

It is possible to construct an extinction--distance relation for the direction towards the pulsar
utilizing  red-clump stars as standard candles, 
following a method described, for instance, 
in \citet{lopez2002AsAp} and \citet{cabrera2005AsAp}.  
The method was used previously to constrain distances and extinctions for several sources. Some examples are 
 the X-ray binary 4U 1608$-$52 \citep{guver2010ApJ},  six anomalous X-ray pulsars   \citep{2006Durant},   
and  two $\gamma$-ray pulsars \citep{danilenko2012AsAp,danilenko2013AsAp}. 

In the top panel of Figure~\ref{fig:nh_dist}, we show $K$ vs $J-K$ band color-magnitude diagram for stars
from the 2MASS All-Sky Point Source Catalogue\footnote{see http://irsa.ipac.caltech.edu/applications/DataTag/, 
DataTag $=$ ADS/IRSA.Gator\#2014/0814/100517\_7624} located within 0\fdg3 of the pulsar position.
The red-clump (RC) branch, as well as main-sequence (MS) and asymptotic-giant (AGB) branches, are indicated. 
We divided the diagram into several magnitude bins, and in each bin 
we fitted $J-K$ color distribution with a mixture of two Gaussians 
corresponding to the MS and RC branches. The AGB stars were eliminated by  omitting 
all points  located right of a boundary starting at $J-K=1.7$ for small magnitudes 
 and ending at $2\fm5$ for large ones. 
The derived $J-K$ colors of RC stars with their uncertainties were
then transformed  into distances and interstellar extinctions $A_V$, using relations from \citet{rieke1985ApJ},  
assuming the absolute magnitude and the intrinsic color of the red-clump stars 
to be $M_{K}$ = $-$1.62~$\pm$~0.03, 
and ($J-K$)$_{0}$~= 0.68~$\pm$~0.07, respectively (see the above-cited papers for details). 
Extinctions  were transformed to  hydrogen absorbing column densities $N_{\rm H}$ using a standard empirical relation
$N_{\rm H}$~= $A_V$~$\times$~(1.79~$\pm$~0.03)~$\times$~10$^{21}$ cm$^{-2}$ \citep{predehl1995AsAp}.

The resulting $N_{\rm H}$--distance dependence is shown in the bottom panel of Figure~\ref{fig:nh_dist}.
$N_{\rm H}$  increases with the distance  reaching  
a limit  of $(15\pm4)\times 10^{21}$ cm$^{-2}$ 
at  distances $\gtrapprox$  5 kpc. 
Within uncertainties this limit is consistent with the total Galactic  $N_{\rm H}$ in the pulsar
direction of $\approx12\times 10^{21}$ and $\approx9.7\times 10^{21}$ cm$^{-2}$,  estimated from the HI maps  
provided by  \citet{DL1990} and \citet{kalberla2005}, respectively. Corresponding $A_V=8.4\pm2.2$ is also compatible with the entire 
Galactic extinction estimate of $\sim$11\fm0 \citep{2011schlafly}, although the respective extinction map is considered 
as not reliable at $b\la5\degs$.

\subsection{X-ray spectral analysis}
\label{sec:x-ray}

We reanalyzed the archival \textit{Chandra} data in light of
the $N_{\rm H}$--distance relation. We extracted the pulsar spectra from all three \textit{Chandra}/ACIS-S 
sets  using  an aperture with a radius of 0\farcs74
centered at the pulsar position applying the \texttt{CIAO} 
\texttt{v.4.6} {\sl specextract} tool. 
We also extracted the PWN spectrum  from an elliptical 
region with semi-axes  of 6\farcs2 and 10\farcs6 and a position angle
of 137\degs, that encloses most of the PWN equatorial torus emission. 
The circle aperture of a 2\arcsec~radius   around the pulsar was excluded from this region. 
Backgrounds were taken from  regions free of any sources on the ACIS-S3 chip, where 
the Dragonfly was exposed in all three \textit{Chandra} data sets with  live times of 19, 59, and 43 ks.
Total numbers of source counts are $\approx$ 1270 for the pulsar and $\approx$ 5250 
for the PWN. 

To evaluate likelihoods, we used the $\chi^{2}$ statistics. To model the pulsar spectrum, we applied an  
absorbed sum of the power-law (PL) and thermal components. Any single component  did not describe the data.   
For the thermal component, we tried blackbody (BB) 
and magnetic neutron star hydrogen atmosphere models 
NSA \citep{pavlov1995lns} and NSMAX \citep{ho2008ApJS}.
For the interstellar absorption, we used the \texttt{XSPEC} photoelectric absorption 
\texttt{phabs} model with 
default cross-sections \texttt{bcmc} \citep{bcmc1992ApJ} 
and abundances \texttt{angr} \citep{angr1989GeCoA}.

To model the contribution of the PWN to the spectrum extracted from the pulsar aperture, we 
added second PL to the pulsar spectral model 
and fitted the PSR and PWN spectra simultaneously in the 0.3--10 keV spectral range. 
The second PL component 
photon index was tied with the PWN photon index, and $N_{\rm H}$ 
was set as a global parameter.
Doing this, we also took into account the ratio of the PWN flux within the pulsar aperture to the total PWN flux of 
$\approx$ 0.05, as it was estimated  by \citet{vanetten2008ApJ} and independently confirmed 
by us via modeling of \textit{Chandra}/ACIS PSF.

The $N_{\rm H}$--distance relation and its uncertainty were approximated 
by smoothing splines, shown by the line and gray filled region in the bottom panel of Figure~\ref{fig:nh_dist}. 
This relation was then used as a Bayesian prior information for the subsequent spectral fitting procedure 
\citep[see, e.g.,][for details]{GelmanBook}.
Technically, we assumed that for each distance the $N_{\rm H}$ value  
follows a Gaussian distribution with the mean 
and $\sigma$ taken from 
the approximations.
We then run Markov chain Monte Carlo (MCMC) 
using the Goodman-Weare algorithm  
implemented as a \texttt{python} 
package \texttt{emcee} by \citet{foreman-mackey2013PASP}.
For each model we kept 1000 steps after initial burn-in, which 
is large enough considering that typical autocorrelation time 
\citep[see, e.g.][]{goodman2010CAMCS} was of order of 
several tens (50--90) of iterations. 
As 100 MCMC walkers \citep{goodman2010CAMCS} was used, we obtained 100000 samples in total.

Posterior median values of spectral parameters with 90\% credible intervals 
for the BB+PL and NSMAX+PL models are presented in Table~\ref{t:x-fit}.
The goodness-of-fit test ($\chi^{2}$ values are  in  Table~\ref{t:x-fit}) 
shows that both models are equally consistent with the data.
We present fit results for only one of hydrogen atmosphere models, NSMAX 1260,
which corresponds to the surface magnetic field B~= 4~$\times$~10$^{12}$ G
\citep{ho2008ApJS}.
The thermal component can be equally well fitted by NSA or by any other model from the NSMAX set;     
the resulting parameters do not 
depend appreciably on the  choice of  specific atmosphere model. We prefer more modern NSMAX models because 
they account for the partial ionization in stellar atmospheres, especially important at low temperatures (as in our case),
while
the NSA models were constructed for fully ionized NS atmospheres.

Our results are generally consistent with those 
presented by \citet{hessels2004ApJ} and \citet{vanetten2008ApJ}.
However, in contrast to \citet{vanetten2008ApJ}, we did not fix $N_{\rm H}$  
at a best-fit value obtained from a separate analysis of the PWN spectrum. 
Nevertheless, the resulting $N_{\rm H}$ is defined mainly by the PWN spectrum, and consistent with the one obtained by \citet{vanetten2008ApJ} 
within uncertainties. Thereby it weakly  depends on a particular model used to describe the pulsar spectrum 
and on a particular $N_{\rm H}$--$D$ relation.
The distance $D= 
1.8^{+1.7}_{-1.4}$ kpc\footnote{Here and below we discuss the largest $D$ range 
from Table~\ref{t:x-fit}.} 
is now mainly determined by $N_{\rm H}$ and the adopted $N_{\rm H}$--$D$ relation. 
In this approach, only two parameters are defined by the thermal component: the  
radius $R$ and the effective temperature $T$ of the emitting area. 
Importantly, we were able to infer the absolute value of the radius $R$,
not only the $R/D$ ratio as it would have been without accounting for the $N_{\rm H}-D$ relation.
As it is typical for  pulsars where X-ray spectral data can be equally well 
fitted by the blackbody and NS atmosphere models \citep[e.g.,][]{2001pavlov, 2014kir}, for the BB model 
$R$ is a factor of 10 smaller and $T$ is a factor of 2.5  larger than those %these values 
for the hydrogen atmosphere model.
For the latter, 
$R$~= 12$^{+20}_{-10}$ km implies that emission can come from the bulk of the NS
surface with the effective surface temperature, redshifted for distant observer, $T$~= 63$^{+9}_{-8}$ eV, 
close to that of the Vela pulsar \citep{2001pavlov}.
For the BB model, 
$R$~= 1.3$^{+1.5}_{-1.0}$ km, which is compatible with a canonical pulsar hot polar cap radius 
of $\sim$ 0.6 km  for a 100-ms pulsar 
\citep{sturrock1971ApJ}. 

\begin{landscape}
\begin{table*}[th]
  \caption{
  Posterior medians for the pulsar and the PWN equatorial 
  torus region spectra.   The BB+PL and NSMAX+PL models are for the pulsar spectrum, while the PWN is described by 
  the  PL model \dag.
  }
  %\begin{center}
  \begin{center}
  \begin{tabular}{cccccccccc}
  %\hline
  \hline
  \multicolumn{6}{c}{ }  \\
Model    & $N_{\rm H}$              & $\Gamma_{\rm psr}$ & $K^{\rm psr}$ \ddag                 &  $T$ &  $R$  & $D$    & $\Gamma_{\rm pwn}$ & $K^{\rm pwn}$ \ddag  & $\chi^{2}$ \\
         & 10$^{21}$ cm$^{-2}$ &          & 10$^{-5}$ photons                       &   eV  &  km    & kpc     &                    & 10$^{-5}$ photons   &     (dof)      \\
         &                     &          & keV$^{-1}$~cm$^{-2}$~s$^{-1}$    &       &        &         &                    & keV$^{-1}$~cm$^{-2}$~s$^{-1}$ &            \\  
         \\
    \hline
   
    \multicolumn{6}{c}{ }  \\

BB+PL    & 5.8$^{+0.5}_{-0.5}$ & 1.8$^{+0.6}_{-0.6}$ & 1.0$^{+1.0}_{-0.6}$   & 155$^{+14}_{-14}$ & 1.3$^{+1.5}_{-1.0}$   & 1.8$^{+1.5}_{-1.4}$  & 1.4$^{+0.1}_{-0.1}$ & 9.7$^{+1.0}_{-0.9}$ & 459(469) \\
\multicolumn{6}{c}{ }  \\

NSMAX+PL & 6.0$^{+0.5}_{-0.5}$ & 1.3$^{+0.7}_{-0.8}$ & 0.5$^{+0.7}_{-0.4}$   & 63$^{+9}_{-8}$    & 12.0$^{+19.5}_{-9.6}$ & 1.8$^{+1.7}_{-1.4}$  & 1.4$^{+0.1}_{-0.1}$ & 10.0$^{+1.0}_{-0.9}$ & 472(469)\\

\multicolumn{6}{c}{ }  \\

\hline
\end{tabular}
\tablenotetext{\dag}{$N_{\rm H}$ is the absorbing column density. The temperatures  $T$ and emitting area radii $R$ for the BB and 
NSMAX spectral components are given as measured by a distant observer. For the NSMAX component, the gravitation 
redshift $1+z=(1-2.952 M/R)^{-0.5}$, where M and R are the NS mass and circumferential  radius in the Solar mass 
and km units, respectively, is fixed at 1.21. This corresponds to a canonical NS with  $M=1.4$ and $R=13$ km. 
$K^{\rm psr,pwn}$ and $\Gamma_{\rm psr,pwn}$ are PL normalizations and photon spectral indexes for the pulsar (psr) and PWN (pwn), 
respectively.     
All errors correspond to 90\% credible intervals derived via MCMC.}
\tablenotetext{\ddag}{Pulsar fluxes in 2--10 keV range are $\log F_{\rm X}^{\rm psr}$ = $-$13.5$^{+0.7}_{-0.8}$, $-$13.4$^{+1.0}_{-1.2}$ [ erg~cm$^{-2}$~s$^{-1}$ ],  for BB+PL and NSMAX+PL models, 
respectively. PWN flux in the same range is $\log F_{\rm X}^{\rm pwn}$ = $-$12.2$^{+0.1}_{-0.1}$ [ erg~cm$^{-2}$~s$^{-1}$ ] for both models.}
\end{center}
\label{t:x-fit}
\end{table*}
\end{landscape}

\subsection{Multiwavelength spectra of the pulsar and PWN}
\label{sec:mws}
\begin{figure}
\setlength{\unitlength}{1mm}
\begin{center}
\begin{picture}(160,165)(0,0)
\put (20,85) {\includegraphics[scale=0.49, clip]{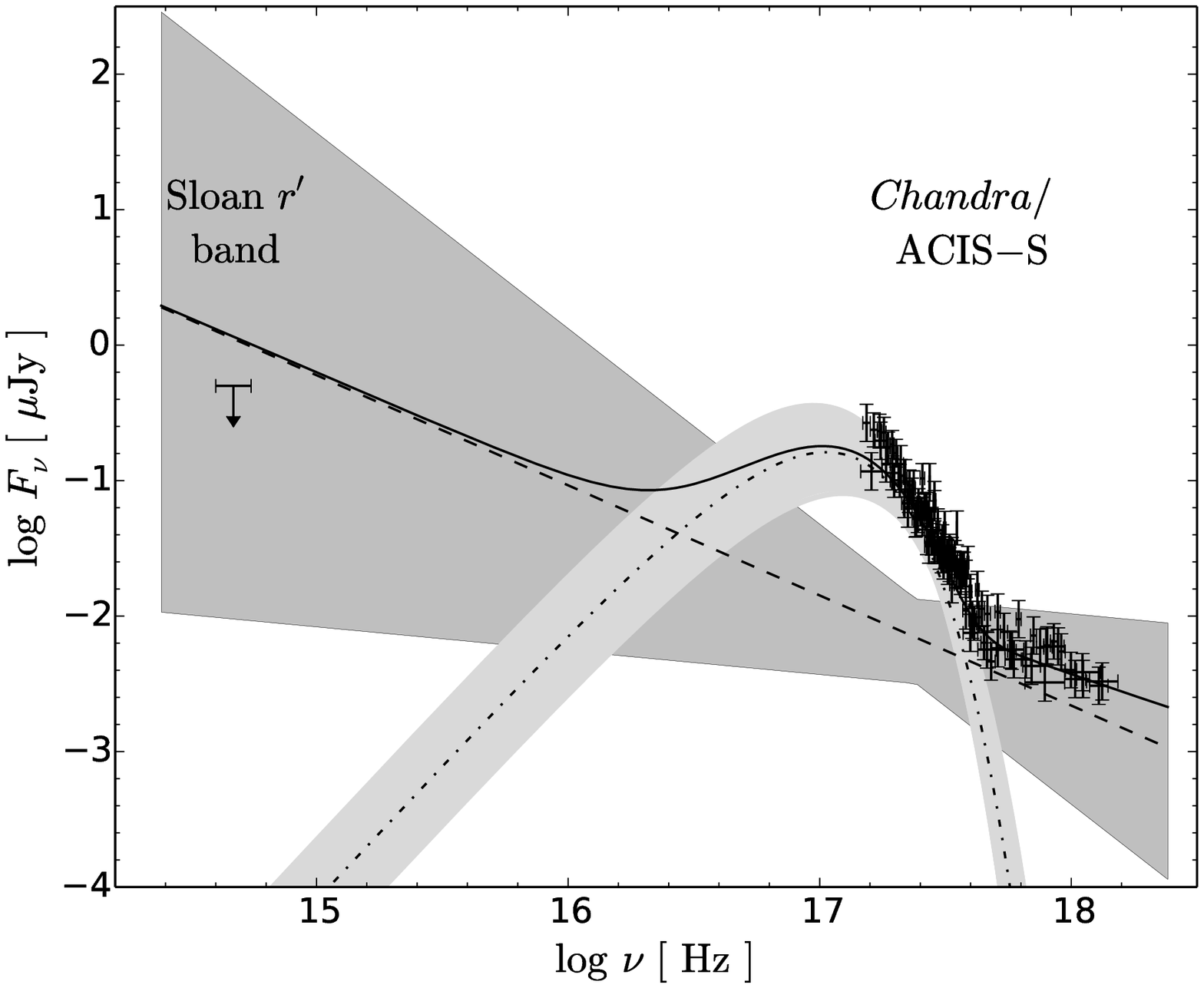}}
\put (20,0)   {\includegraphics[scale=0.49, clip]{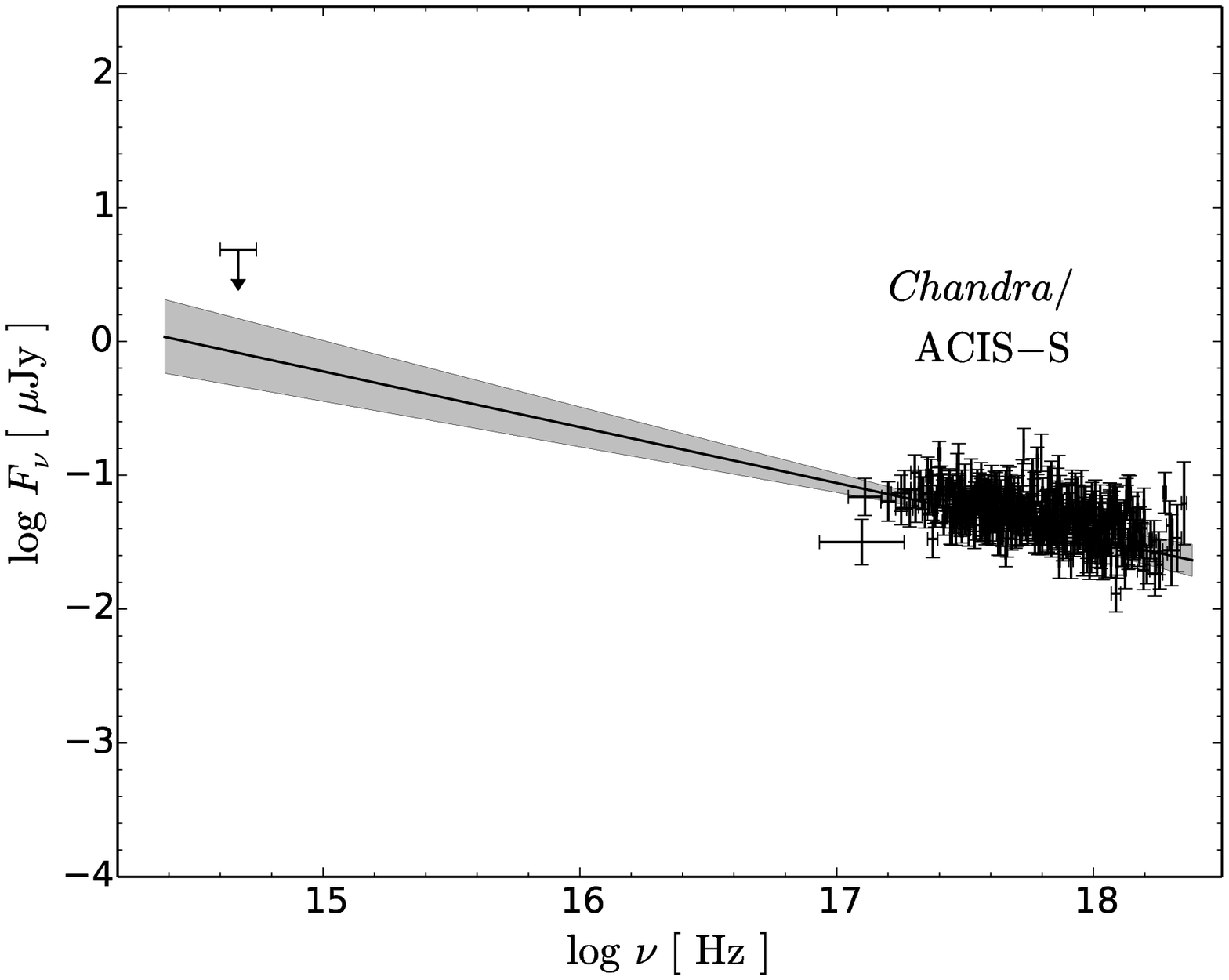}}
\end{picture}
\end{center}
 \caption{
 {\sl Top}: Unabsorbed spectrum of \psr. The solid line is the best-fit model for the  \textit{Chandra} 
 X-ray spectrum which includes BB and PL pulsar spectral components and the PWN contribution to 
 the spectral extraction aperture (see text for details). The best-fit model is extrapolated towards the 
 optical. The GTC dereddened  3$\sigma$ flux upper limit in the $r'$ band is shown by the bar with the arrow. 
 Dash-dotted and dashed lines with light- and dark-gray regions are the BB and PL pulsar spectral components 
 with their 90\% uncertainties, respectively. The difference between the solid and dashed lines  is 
 clearly visible at the high-energy tail and reflects the PWN contribution.
 {\sl Bottom}: The solid line  and gray region are the best-fit PL model of the X-ray spectrum of the PWN equatorial 
 torus region with its 90\% uncertainties, respectively.
 The GTC dereddened optical 3$\sigma$ flux upper limit for this region in the $r'$ band is also shown. 
 The error-bar crosses in each panel are the unfolded \textit{Chandra} data.
 }
 \label{fig:spectrum}
\end{figure}

The best-fit $N_{\rm H}$ values obtained from the X-ray spectral analysis suggest  
a total interstellar extinction towards the Dragonfly $A_V\approx3.3$  in the $V$ band, which results in the extinction  
$A_{r'}\approx2.8$ 
in the $r'$ band using a standard extinction law with $R_V=3.1$ \citep{1989cardelli}.  
Based on this,   upper limits on the dereddened flux densities   for the pulsar 
and PWN in the $r'$ band are about 0.57 $\mu$Jy  and  4.85 $\mu$Jy, respectively. 
In Figure~\ref{fig:spectrum} we compare these limits  with unabsorbed X-ray spectra of the pulsar (top panel) 
and PWN (bottom panel), fitted by 
BB+PL and PL models, respectively. 
For the PWN, the optical and X-ray data were obtained from the same spatial region enclosing its 
X-ray equatorial torus emission (see Sect.~\ref{sec:opt_res}, \ref{sec:x-ray}, and Figure~\ref{fig:ima}). 
The solid line in the top panel of Figure~\ref{fig:spectrum}  
shows the total best-fit model, including the contribution of PWN nonthermal photons to 
the  spectrum extracted from the pulsar aperture.
The dashed line shows solely the PL component of the pulsar.
As seen, the PWN contribution is substantial  only in the high-energy tail.     

As seen from Figure~\ref{fig:spectrum}, the pulsar optical 
flux upper limit does not exceed the  
extrapolation of the best-fit
X-ray spectral model to the optical.
This is typical for rotation powered pulsars detected in the optical and X-rays 
\citep{danilenko2011MNRAS,2011durant,2010mig,2006shib}. 
For all of them, the nonthermal component dominates in the optical. However, 
this component usually  shows a break between the optical and X-rays 
with a significant  spectral flattening in the optical. Our data do not exclude the presence of 
such a  break for \psr, although  the  extrapolation  of the X-ray PL component is still rather uncertain 
and the optical limit is not deep enough.
The NSMAX+PL model, which equally well fits the X-ray data, does not change these conclusions.  

All torus-like PWNe, which have been detected in both spectral domains, also show spectral flattening in the optical 
in comparison with X-rays 
\citep[e.g.,][]{zharikov2013ApJ}. However,  
for the Dragonfly the situation  is currently even less certain than for its pulsar. The nebula optical 
flux upper limit overshoots the low-energy extrapolation of its X-ray spectrum (bottom panel of Figure~\ref{fig:spectrum}), 
and the presence of the break in the spectrum of this PWN remains an open question.     

The dereddened upper limits were obtained using the $A_V-N_{\rm H}$ relation of \citet{predehl1995AsAp}. 
There exist other empirical relations of that kind. For instance, \cite{GuverOzel2009}
give $N_{\rm H}$~= $A_V$~$\times$~(2.21~$\pm$~0.09)~$\times$~10$^{21}$ cm$^{-2}$, also consistent with the results of
\citet{Gorenstein1975}. Using this relation instead, we get smaller $A_V \approx$ 2.7 and hence  dereddened
upper limits a smaller by a factor of 1.6, which does not change general conclusions of this Section.

%%%%%%%%%%%%%%%%%%%%% dist_diag  %%%%%%%%%%%%%%%%%%%%%%%%%%%%%%%%%%%%%%%
%%%%%%%%%%%%%%%%%%%%%%%%%%%%%%%%%%%%%%%%%%%%%%%%%%%%%%%%%%%%%

\section{Dicussion}
\label{sec:dis}

Our rather deep, down to $r'\approx27.2$, GTC optical imaging of the Dragonfly Nebula field allowed 
us to set  upper limits on the optical flux densities of \psr\ and its PWN.
The  non-detection of this  
energetic system  can be attributed   to  high interstellar extinction towards the object, which is 
roughly about of 3$^m$ in the $r'$ band.   
Considering the data for other pulsars detected in the optical and X-rays, 
we conclude that 1--2 magnitude deeper optical observations
are necessary to detect this 
pulsar  and to reveal  the  
expected spectral break between the optical and X-rays. 
This is, in principle, feasible with  8-10 m ground-based telescopes, such as the GTC,  
using a few hour exposure in $r'$ at good seeing conditions,  
although observations at longer wavelengths, less affected by the interstellar absorption,   
would be preferable. Such observations would also be  useful  to better constrain the optical-X-ray 
spectral properties of the PWN.

Our $N_{\rm H}$--distance relation, constructed using the red-clump star method and 
compiled with the X-ray spectral analysis, supports previous suggestions  that 
the pulsar is likely to be substantially closer to us than it is inferred from  DM  and the NE2001 model of 
the  Galactic distribution of free electrons.  Our estimate  
$D$~= 1.8$^{+1.7}_{-1.4}$~kpc  
is compatible, within uncertainties, with the 3--4 kpc range suggested by \citet{vanetten2008ApJ}. 
However, our allowed distance range is shifted to lower distances.  It 
suggests the association of the pulsar 
with the Cygnus-X region, located within 2 kpc from the Sun,  and is consistent with the $\gamma$-ray 
``pseudo-distance''  of $\sim1$ kpc provided by 
the \textit{Fermi} data. 
The reduced distance we found makes feasible parallax and pulsar proper motion measurements with VLBI. 
A possible source of systematic errors in our distance determination method 
originates from ambiguity in $A_V-N_{\rm H}$ relations, as stated above. Reprocessing the analysis of 
Sections~\ref{sec:dist_ext} and \ref{sec:x-ray} using the relation of \citet{GuverOzel2009}, we obtain 
an even smaller distance of $1.3^{+1.5}_{-1.1}$~kpc.

Comparing the DM of 370 pc~cm$^{-3}$, or electron column density $N_e$ $\approx$ 1.14~$\times$~10$^{21}$ cm$^{-2}$, 
with $N_{\rm H}$ of 6$\times 10^{21}$ cm$^{-2}$ 
leads to an average ionization ratio of 19\% along the pulsar line of sight
which is not too much larger
than the 10\% ionization found on average \citep[e.g.,][figure~1]{HeNgKaspi2013}.
On the other hand, the NE2001 electron density model in the pulsar direction 
gives much smaller $N_e$ $= 0.7^{+1.2}_{-0.6}\times$~10$^{20}$ cm$^{-2}$ for
distance range of $D$~= 1.8$^{+1.7}_{-1.4}$~kpc. There exist several indications that the NE2001 
model strongly underestimates $N_e$ in the vicinity of the \psr\ direction \citep{Camilo2009ApJ,Camilo2012ApJ,arumugasamy2014ApJ}. 
For instance, there is another 
pulsar J2022+3842 at only 1\fdg8 from \psr\ with very high DM~= 429 pc cm$^{-2}$, for which NE2001 model gives 
obviously overestimated distance $D$~$>$ 50 kpc 
\citep{arumugasamy2014ApJ}. This may imply that there are dense clouds in the Cygnus-X region, 
which are not taken into account in the NE2001 model \citep{roberts2002ApJ}. 
 
As for the X-ray thermal emission component of \psr, we cannot state definitely whether the thermal emission 
comes from 
a hot polar cap or the bulk of the NS surface. This depends 
on which, BB or a hydrogen atmosphere, model is applied to describe the 
thermal emission. 
The phase-resolved spectroscopy would be useful to distinguish between the two  
possibilities. 
We  note, however, that if the thermal emission originates from the entire surface of  
the star (the case of the atmospheric model), \psr\ has a rather small surface 
temperature for its age. According to the NS cooling theories, such 
a small temperature can not be explained by a standard cooling scenario, 
but it can be reached if the effects of superfluidity in the stellar 
interiors are invoked. This is possible, for instance,  
if the powerful direct Urca process of neutrino emission operates in 
the star but is suppressed by superfluidity \citep{yakovlev2004ARA}. 
In  another interpretation the direct Urca process is not allowed, 
but the cooling is enhanced due to the specific process of the neutrino 
emission accompanying the Cooper pair formation in the neutron triplet superfluid. 
This is the essence of the so-called minimal cooling scenario \citep{gusakov2004AsAp,page2004ApJS,page2009ApJ}. 

%%%%%%%%%%%%%%%%%%%%%%%%%%%%%%%%%%%%%%%%%%%%%%%%
\begin{figure*}[t]
\setlength{\unitlength}{1mm}
\begin{center}
\begin{picture}(160,80)(0,0)
\put (0,80) {\includegraphics[width=78.0mm, angle=-90, clip=]{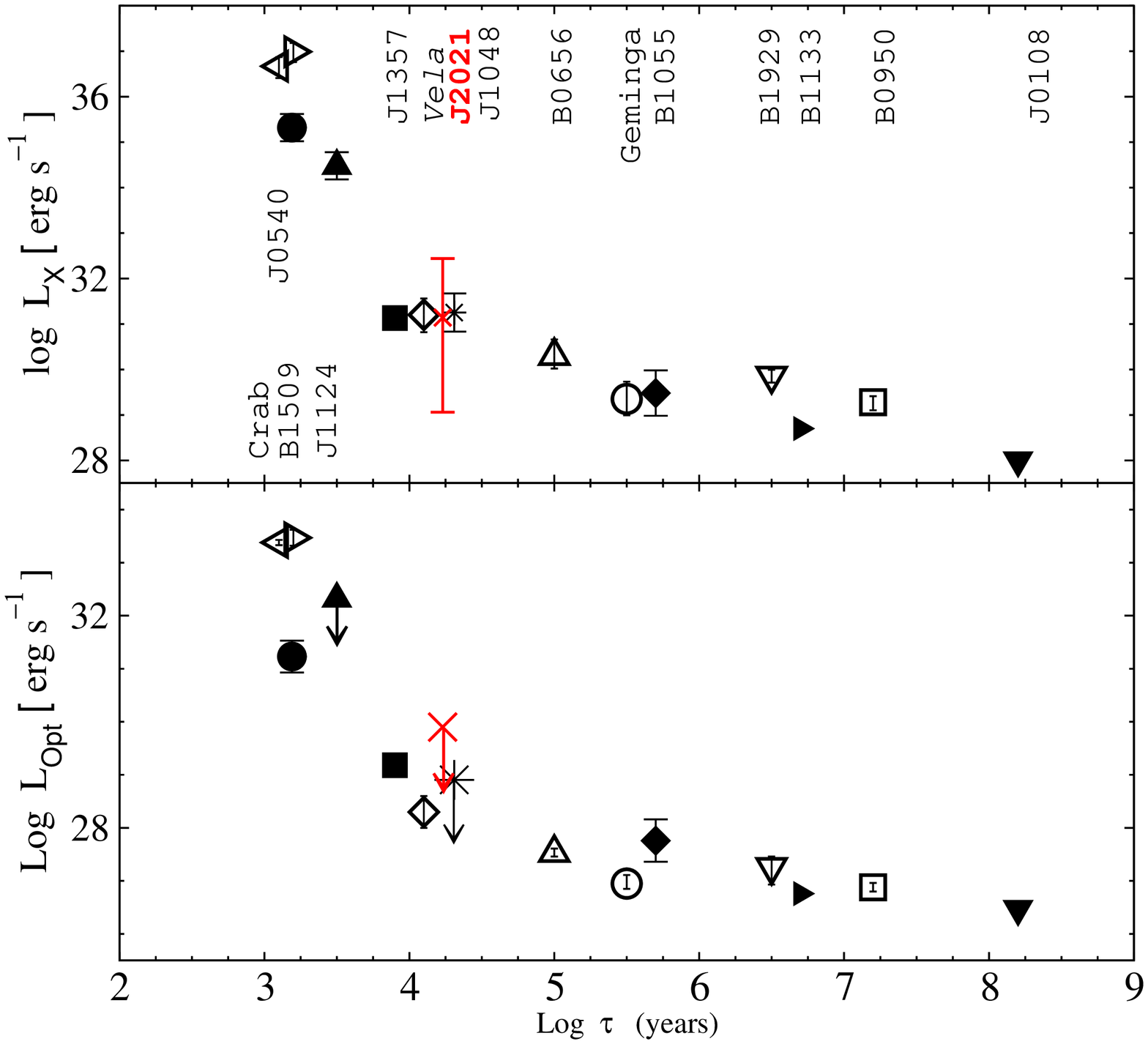}}
\put (80,80) {\includegraphics[width=78.0mm,angle=-90, clip=]{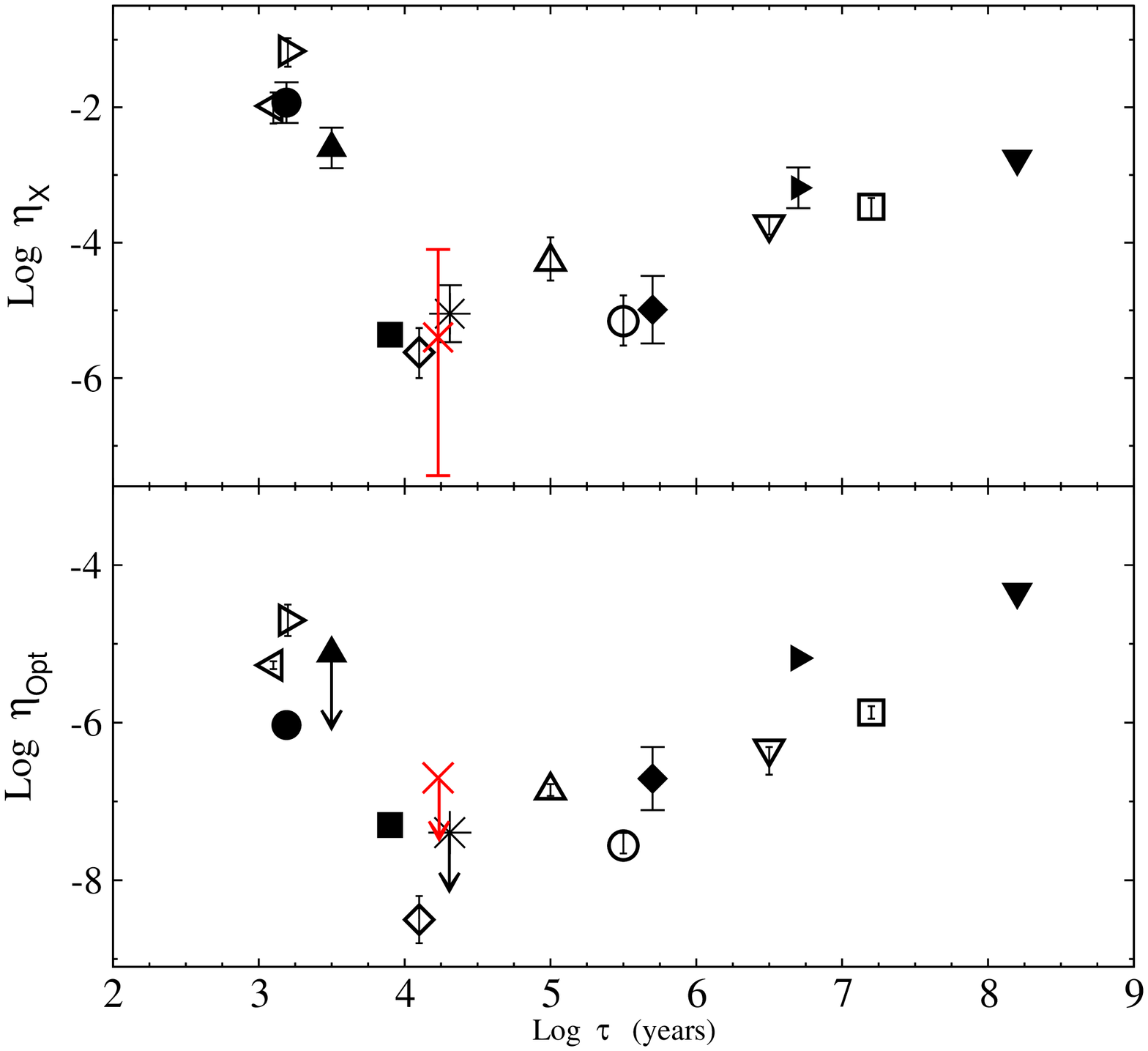}}
\end{picture}
\end{center}
%\vspace{-57mm}
%\hfill\parbox[b]{80mm}{
\caption{
X-ray and optical luminosities, $L_{\rm X}$ and $L_{\rm Opt}$, and respective efficiencies, $\eta_X$ and $\eta_{\rm Opt}$, for pulsars 
of different characteristic age $\tau$ detected in both spectral domains. The data are adopted from \citet{danilenko2013AsAp}. 
Different pulsars are marked by different symbols. The  \psr\ data, derived in this work, are included (marked by asterix). 
 }
% }
%\vspace{10mm}
\label{fig:lum-eff}
\end{figure*}
%%%%%%%%%%%%%%%%%%%%%%%%%%%%%%%%%%%%%%%%%%%%%%%%%%%%
The pulsar's 0.1--100 GeV  $\gamma$-ray luminosity   $L_{\gamma}\approx 5.9\times10^{36}$ erg s$^{-1}$
and efficiency  $\eta_{\gamma}=L_{\gamma}/\dot{E}\approx1.8$, derived in the 2nd Fermi Pulsar Catalog 
\citep{2013Abdo} using the distance  of 10 kpc from \citet{hessels2004ApJ} appear to be
unreasonably high, and 
place the pulsar at the highest end of $L_{\gamma}$ and $\eta_{\gamma}$ distributions 
of $\gamma$-ray pulsars.      
In contrast, for the distance $D$~= 1.8$^{+1.7}_{-1.4}$ kpc inferred 
from our analysis, $\log L_{\gamma}$~= 35.3$^{+0.6}_{-1.3}$ [ erg s$^{-1}$ ]
and $\log \eta_{\gamma}$~= $-1.2^{+0.6}_{-1.3}$
become consistent with the average values of the respective distributions 
for $\gamma$-ray pulsars with similar $\dot{E}$ and/or characteristic age 
\citep[cf., Figures~9 and 10 of][]{2013Abdo}.   

The pulsar's unabsorbed nonthermal X-ray flux in 2--10 keV range, derived from the X-ray spectral fits, is 
$\log F_{\rm X}$~= $-$13.5$^{+0.7}_{-0.8}$ [ erg cm$^{-2}$ s$^{-1}$ ]. 
The respective X-ray luminosity and efficiency are 
$\log L_{\rm X}$~= 31.1$^{+1.3}_{-2.1}$ [ erg s$^{-1}$ ] and $\log \eta_{\rm X}$~= $-$5.4$^{+1.3}_{-2.0}$,
assuming the distance  range
derived in this work. 

For the 90\% distance upper limit $D=3.5$ kpc, 
upper limits on 
the optical luminosity and efficiency in the $V$ band, assuming a flat spectrum, 
are $\log L_{\rm Opt}$~$\lesssim$ 29.9 [ erg~s$^{-1}$ ] 
and $\log \eta_{\rm Opt}\la -6.7$, respectively. 
In Figure~\ref{fig:lum-eff} we compare the obtained X-ray and optical 
efficiencies and luminosities with the data for  other pulsars observed in both ranges  \citep{danilenko2013AsAp}.
According to Figure~\ref{fig:lum-eff}, we can conclude that \psr, like the Vela pulsar, is inefficient in these ranges 
as compared to other substantially younger and older pulsars. 
We thus obtain a new member of the small sample of Vela-like pulsars forming  a puzzling minimum   
in the optical and X-ray efficiency  dependences on age  at a characteristic age of $\sim$~10 kyr 
noticed previously by \citet{zhar}.  
No such minimum is visible in the respective $\gamma$-ray dependence \citep{2013Abdo}.  
Together with strong glitches and high polarization in the radio \citep{hessels2004ApJ}, 
a bright NS thermal emission component in X-rays, a double arc X-ray PWN with jets  \citep{vanetten2008ApJ}, 
$\gamma$-ray activity \citep{abdo2009ApJ},  
and  association with a TeV source 
\citep{aliu2014ApJ},
this makes the Dragonfly pulsar and PWN  remarkably similar to the Vela pulsar and its PWN.

\acknowledgments

We thank Dima Barsukov and Serge Balashev for helpful discussion and the anonymous referee for useful comments.
The work was supported by the Russian Science Foundation, grant 14-12-00316.
The scientific results reported in this article are partially based 
on data obtained from the Chandra Data Archive, 
observations made by the Chandra X-ray Observatory.
The research has also made use of the NASA/IPAC Infrared Science Archive, 
which is operated by the Jet Propulsion Laboratory, 
California Institute of Technology, under contract with 
the National Aeronautics and Space Administration.

{\it Facilities:} \facility{Gran Telescopio Canarias, \textit{CXO}}.  %%%, \facility{HST (STIS)}, \facility{CXO (ASIS)}.

%%%%%%%%%%%%%%%%%%% REFERENCES %%%%%%%%%%%%%%%%%%%%%%%%%%%%%%%%%%%%%%%%%%%%%%%%%%%%%%%%
%\bibliographystyle{unsrtnat}
\bibliographystyle{apj}
\bibliography{ref}

%% Use the figure environment and \plotone or \plottwo to include
%% figures and captions in your electronic submission.
%% To embed the sample graphics in
%% the file, uncomment the \plotone, \plottwo, and
%% \includegraphics commands
%%
%% If you need a layout that cannot be achieved with \plotone or
%% \plottwo, you can invoke the graphicx package directly with the
%% \includegraphics command or use \plotfiddle. For more information,
%% please see the tutorial on "Using Electronic Art with AASTeX" in the
%% documentation section at the AASTeX Web site, http://aastex.aas.org/
%%
%% The examples below also include sample markup for submission of
%% supplemental electronic materials. As always, be sure to check
%% the instructions to authors for the journal you are submitting to
%% for specific submissions guidelines as they vary from
%% journal to journal.

%% This example uses \plotone to include an EPS file scaled to
%% 80% of its natural size with \epsscale. Its caption
%% has been written to indicate that additional figure parts will be
%% available in the electronic journal.

\clearpage

%% Here we use \plottwo to present two versions of the same figure,
%% one in black and white for print the other in RGB color
%% for online presentation. Note that the caption indicates
%% that a color version of the figure will be available online.
%%

\end{document}